\documentclass[twocolumn,twocolappendix,apjl,appendixfloats]{aastex63}
\usepackage{amsmath}
\usepackage{amssymb}

\newcommand{\p}{\partial}
\usepackage{graphics}
\usepackage{graphicx}
\usepackage{footnote}
\usepackage{txfonts}
\usepackage{float}
\if 0
\usepackage{titlesec}

\titlespacing*{\section}
{0pt}{1ex }{1ex}
\titlespacing*{\subsection}
{0pt}{1ex }{1ex}
\vspace{-0.3cm}

\fi 
\makesavenoteenv{tabular}

\DeclareSymbolFont{cmletters}{OML}{cmm}{m}{it}
\DeclareMathSymbol{v}{\mathalpha}{cmletters}{"76}

\usepackage{hyperref}
\definecolor{darkblue}{rgb}{0.0,0.0,0.3}
\hypersetup{colorlinks,breaklinks,
            linkcolor=darkblue,urlcolor=darkblue,
            anchorcolor=darkblue,citecolor=darkblue}

\shorttitle{Ab Initio Arrested Development}
\shortauthors{Ressler, White, Quataert, and Stone}

\begin{document}
  
  % Title and author information
\title{Ab Initio Horizon-Scale Simulations of Magnetically Arrested Accretion in Sagittarius A* Fed by Stellar Winds}
\author[0000-0003-0220-5723]{Sean~M.~Ressler}
\affiliation{Kavli Institute for Theoretical Physics, University of California Santa Barbara, Kohn Hall, Santa Barbara, CA 93107, USA}
\author{Christopher~J.~White}
\affiliation{Kavli Institute for Theoretical Physics, University of California Santa Barbara, Kohn Hall, Santa Barbara, CA 93107, USA}
\author[0000-0001-9185-5044]{Eliot~Quataert}
\affiliation{Department of Astronomy, University of California Berkeley, 501 Campbell Hall, Berkeley, CA 94720, USA}
\author{James~M.~Stone}
\affiliation{Institute for Advance Study, 1 Einstein Drive, Princeton, NJ, 08540, USA}

\begin{abstract}
We present 3D general relativistic magnetohydrodynamic (GRMHD) simulations of the accretion flow surrounding Sagittarius A* that are initialized using larger-scale MHD simulations of the $\sim$ 30 Wolf--Rayet (WR) stellar winds in the Galactic center.
The properties of the resulting accretion flow on horizon scales are set not by ad hoc initial conditions but by the observationally constrained properties of the WR winds with limited free parameters. 
For this initial study we assume a non-spinning black hole.
Our simulations naturally produce a $\sim 10^{-8} M_\odot$ yr$^{-1}$ accretion rate, consistent with previous phenomenological estimates.  
We find that a magnetically arrested flow is formed by the continuous accretion of coherent magnetic field being fed from large radii.  
Near the event horizon, the magnetic field is so strong that it tilts the gas with respect to the initial angular momentum and concentrates the originally quasi-spherical flow to a narrow disk-like structure.
We also present 230 GHz images calculated from our simulations where the inclination angle and physical accretion rate are not free parameters but are determined by the properties of the WR stellar winds. 
The image morphology is highly time variable.  
Linear polarization on horizon scales is coherent with weak internal Faraday rotation.

\end{abstract}

%\begin{keywords}
%Galaxy: center -- accretion, accretion disks -- magnetohydrodynamics (MHD) -- stars: Wolf--Rayet  -- black hole physics
%\end{keywords}
\section{Introduction}
Sagittarius A* (Sgr A*), the $\sim 4 \times 10^6 M_\odot$  \citep{GravityS2,Do2019} black hole in the center of our Galaxy, is perhaps the most important low-luminosity active galactic nucleus for testing our understanding of accretion models.  This is in part because we have a clear picture of how the accretion flow is fed via the powerful stellar winds of the $\sim$ 30 Wolf--Rayet (WR) stars orbiting the black hole \citep{Paumard2006}.  The wind speeds, mass-loss rates, and orbits are well constrained by infrared \citep{Martins2007} and radio observations \citep{YZ2015}, with $\sim$ half of the WR stars confined to a relatively thin clockwise stellar disk \citep{Belo2006,Lu2009}.  These winds can account for a majority of the accretion budget of Sgr A*. This view is corroborated by semi-analytic models and 3D simulations of wind-fed accretion that produce accretion rates, X-ray luminosities, and even rotation measures that are consistent with the observed values/constraints \citep{Quataert2004,Cuadra2008,Shcherbakov2010,Russell2017,Ressler2018,Ressler2019,Ressler2020,Calderon2020}.  

Given this knowledge of how the black hole is fueled, the Galactic center provides a unique opportunity to determine, from first principles, the state of accretion at event horizon scales by calculating how the gas provided by the WR stellar winds falls inwards.  This has been the overarching goal of \citet{Ressler2018,Ressler2019,Ressler2020}, hereafter, R18, R19, and R20, respectively, where we presented 3D hydrodynamic and magnetohydrodynamic (MHD) simulations that treat the winds as source terms of mass, momentum, energy, and magnetic field (building on earlier hydrodynamic work by \citealt{Cuadra2005,Cuadra2006,Cuadra2008}).  Both hydrodynamic and MHD simulations displayed similar dynamics, with accretion through the inner boundary proceeding mainly through radial, low angular momentum streams of gas sourced by 1--3 stellar winds, largely confirming the picture first proposed by \citeauthor{Loeb2004} (\citeyear{Loeb2004}, with WR stars replacing the SO stars used in that calculation).  Unfortunately, covering the entire dynamic range of accretion that spans $\sim$ 7 orders of magnitude in radius is impossible in a single simulation because of the large discrepancy in time-scales, so our previous works were only able to reach $\sim 300 r_{\rm g}$ (starting at $\sim $ pc $\approx  5 \times 10^6 r_{\rm g}$ scales), where $r_{\rm g} = M$ is the gravitational radius of the black hole.  Here and throughout we set the gravitational constant and the speed of light to unity, $G=c=1$.   

In this letter, we apply a new technique that allows us to extend the results of our previous simulations to the event horizon in full general relativistic magnetohydrodynamics (GRMHD).  We do this using an intermediate MHD simulation that bridges the gap between large and small scales, essentially resulting in a self-consistent wind-fed GRMHD solution with few free parameters.   All past GRMHD models had the freedom to arbitrarily choose, e.g., the magnetic field geometry and the inclination of the accretion disk with respect to the line of sight while also being able to arbitrarily scale the accretion rate to match observations.  Here we have significantly less freedom, with the properties of the accretion flow being determined by the observationally-constrained stellar winds at large radii.

%This letter is organized as follows. \S 2 briefly describes our computational methods, \S 3 presents our simulation results, while \S 4 discusses these results and concludes. 

\section{Methods}
\label{sec:methods}
All simulations are performed using {\tt Athena++}\footnote{\href{https://princetonuniversity.github.io/athena/}{https://princetonuniversity.github.io/athena/}} \citep{White2016,Athena++}, a conservative, grid-based code for fluid dynamics with mesh refinement, MHD, and GRMHD capabilities.   We use piecewise-linear reconstruction and the Harten--Lax--van Leer+Einfeldt (HLLE, \citealt{Einfeldt1988}) Riemann solver.  The simulation is performed in Cartesian Kerr--Schild (CKS, \citealt{Kerr1963}) coordinates  for a black hole spin $a=0$ via the user-defined coordinate module.  

We generate a realistic, observationally motivated set of initial and boundary conditions for our GRMHD simulations using the wind-fed MHD simulations of R20 by running an intermediate MHD simulation to bridge the gap between large and small scales.  This technique is detailed thoroughly in Appendix \ref{app:inits}, illustrated in Figure \ref{fig:domain}, and demonstrated in Appendix \ref{app:3_sim_results}.  Essentially the only free parameters in the R20 simulations are the ratio between the ram pressure and the magnetic pressure in each WR stellar wind, $\beta_{\rm w}$, and the (randomly chosen) orientation of the spin axes of the stars.   R20 found that the qualitative simulation results were insensitive to the latter, so we focus here on one particular realization of the spin axes for $\beta_{\rm w} = 10^2$ and $\beta_{\rm w} = 10^6$.  

\begin{figure}
\includegraphics[width=0.45\textwidth]{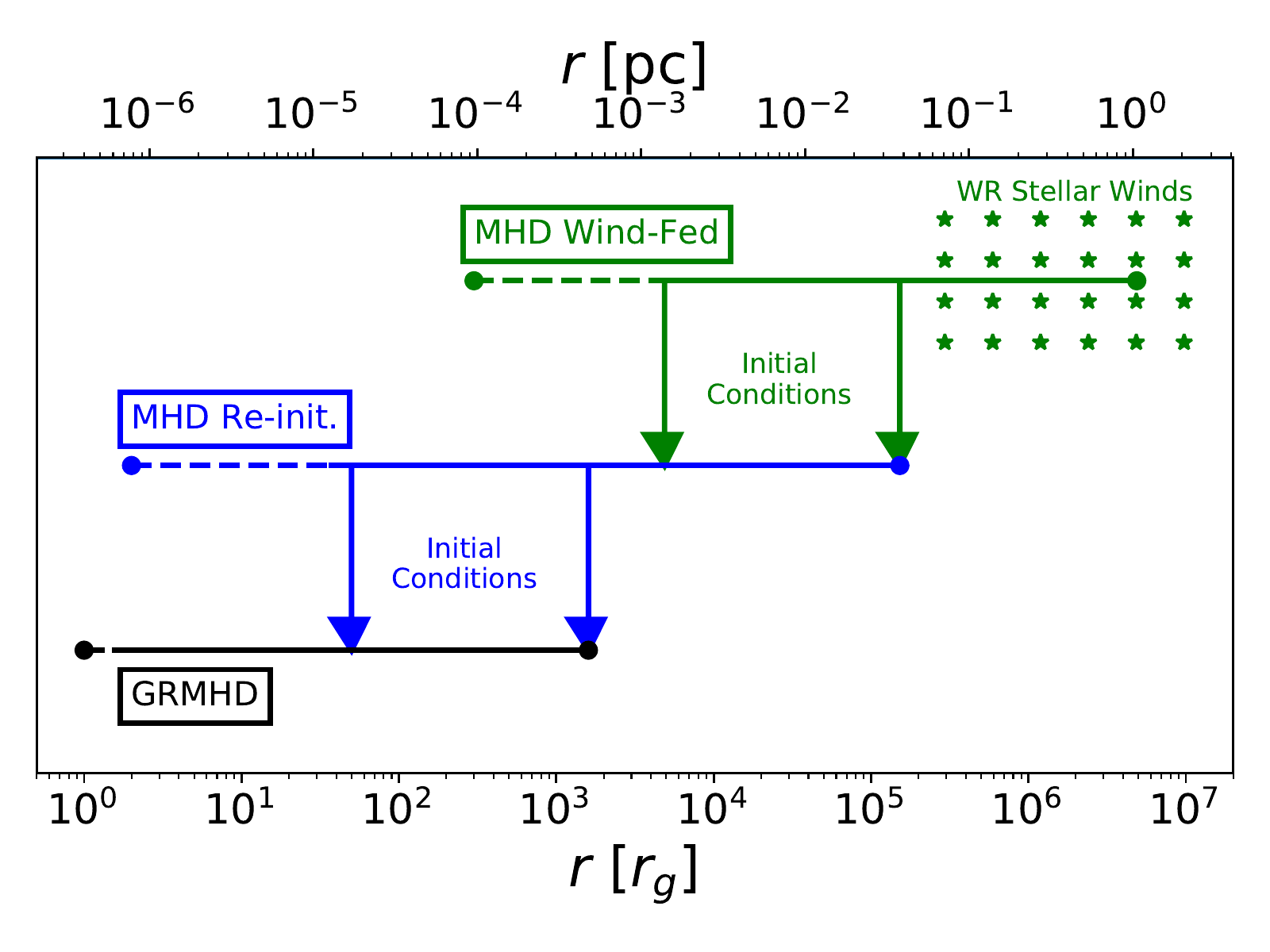}
\caption{Schematic of how we generate initial conditions for GRMHD simulations (black) from the large-scale, MHD simulations of WR stellar wind-fed accretion presented in R20 (green) using an intermediate-scale MHD simulation that is re-initialized using R20 data (blue).   Line segments show the radial domain of each simulation with the dashed portion indicating where the effective logarithmic radial spacing breaks down once the finest level of mesh refinement has been reached, arrows indicate the radial range of simulation data used for initial conditions in the corresponding smaller scale simulation (pointing towards the simulation that received the data), while the asterisks denote the region containing the WR stellar winds.  From top to bottom, the simulations are run for 1.25 kyr, 0.24 yr, and 20,000 $M$ $\approx 5.2$ days for Sgr A*.  } 
\label{fig:domain}
\end{figure}

The simulation domains are (3200 $r_{\rm g}$)$^3$ cubes centered on the black hole with a base resolution of 128$^3$ and 9 levels of nested static mesh refinement (SMR) to mimic logarithmic spacing in radius.  The highest level of refinement is contained within a (6.25 $r_{\rm g}$)$^3$ cube centered on the black hole and has a spacing of $\Delta x _{\rm min} \approx 0.05 r_{g}$.  This ensures that the event horizon is well resolved. Within $r$ = $r_{\rm H}/2$, where $r_{\rm H}$ is the event horizon radius, the density, $\rho$, and pressure, $P$, are set to the numerical floors, the four-velocity is set to free-fall and the magnetic field is allowed to freely evolve (that is, given the floored fluid variables the induction equation is solved without any modification).  This ``inner boundary'' is causally disconnected from everything outside the horizon so it does not affect the solution in the domain of interest.  The density floor is $10^{-6} (r/r_{\rm g})^{-3/2}$ and the pressure floor is $3.33 \times 10^{-9} (r/r_{\rm g})^{-5/2}$, with $\sigma \equiv b^2/\rho \le 100$ and $\beta \ge 0.001$ enforced via additional density and pressure floors, respectively.  Here $\beta$ is the ratio between the thermal and magnetic pressures while $b^2$ is twice the magnetic pressure in Lorentz-Heaviside units.  Additionally, the velocity of the gas is limited such that the maximum Lorentz factor is 50.  The simulations run for 20,000 $M$, a free-fall time at $r \approx 740 r_{\rm g}$.  The adiabatic index of the gas is $\gamma = 5/3$.

For calculating images and polarization we use the publicly available code {\tt grtrans}\footnote{\href{https://github.com/jadexter/grtrans}{https://github.com/jadexter/grtrans}} \citep{Dexter2009,grtrans}, a ray-tracing algorithm that solves the full radiative transfer equation.  Thermal synchrotron emission and absorption are included while inverse Compton scattering is neglected, a good approximation for the 230 GHz frequency we focus on here.  
%In order to make our simulation data  onto a spherical Kerr--Schild grid.   
Since the gas around Sgr A* is essentially a collisionless plasma (e.g., \citealt{Mahadevan1997,Narayan1998}), the electron temperature is not necessarily the same as the total gas temperature given by the GRMHD calculation and must be assigned in post-processing.  We choose to adopt the electron temperature model of \citet{EHT5}, simplified from its more generic form in \citet{Mosci2016}, where the ion to electron temperature is set by the relation $T_i/T_e = (R_{\rm high} \beta^2 + 1)/(1 + \beta^2)$, where $R_{\rm high}$ is the ion to electron temperature ratio for $\beta \gg 1$ and $\beta$ is the ratio between thermal and magnetic pressure.  We use this particular framework because it is easy to implement, widely used in the community, and physically motivated by some calculations of heating by turbulence and magnetic reconnection. That said, there is a great deal of freedom in the electron temperature model, and this $R_{\rm high}$ prescription is only one of many reasonable choices (e.g., \citealt{Mosci2009,CK2015,Anantua2020}).  A more sophisticated treatment of $T_e$ would entail solving the electron entropy equation alongside the GRMHD equations as in \citet{Ressler2015} or \citet{Sadowski2017}, with some physically-motivated model for how dissipation is partitioned between electrons and ions (e.g., \citealt{Howes2010,Rowan2017,Werner2018}, see \citealt{Chael2019,Dexter2020} for discussion).   We are actively exploring other electron temperature models and how they affect the emission predicted by our simulations.  Another important consideration is that a fraction of the electrons are likely accelerated to nonthermal energies by shocks or reconnection (e.g, \citealt{Sironi2011}) and this can have significant consequences for the X-ray, NIR, and low frequency radio emission \citep{Ozel2000,Yuan2003,Ball2016,Chael2017}.  In this work we do not include emission from nonthermal electrons.
Since WR stars typically lack hydrogen \citep{Martins2007}, we calculate the mean molecular weight with no hydrogen and 3 times solar metal abundances.    
%Exploring other electron temperature prescriptions (e.g., \citealt{Anantua2020}), molecular abundances, or dynamically evolving the electron temperature in the simulation itself \citep{Ressler2015,Sadowski2017} is beyond the scope of this work. 

\section{Results}

\subsection{Dynamics}

To facilitate analysis, we define two useful quantities integrated over the horizon: the accretion rate $\dot M$ and the magnetic flux threading the event horizon $\Phi_{\rm BH}$ which is often normalized as $\phi_{\rm BH} \equiv \sqrt{4{\rm \pi}} \textrm{ }\Phi_{\rm BH}/\sqrt{|\dot M|}$ (e.g., \citealt{Sasha2011}). In our Lorentz--Heaviside units, the saturation value for the magnetically arrested (MAD) state is $\phi_{\rm BH} \approx$ 40--60 \citep{Sasha2011,Narayan2012,White2019}, where the MAD state (\citealt{Narayan2003,Igumenshchev2003,Sasha2011}) is one in which the outwards Lorentz force is strong enough to halt the inflow of gas.

\begin{figure}
\includegraphics[width=0.45\textwidth]{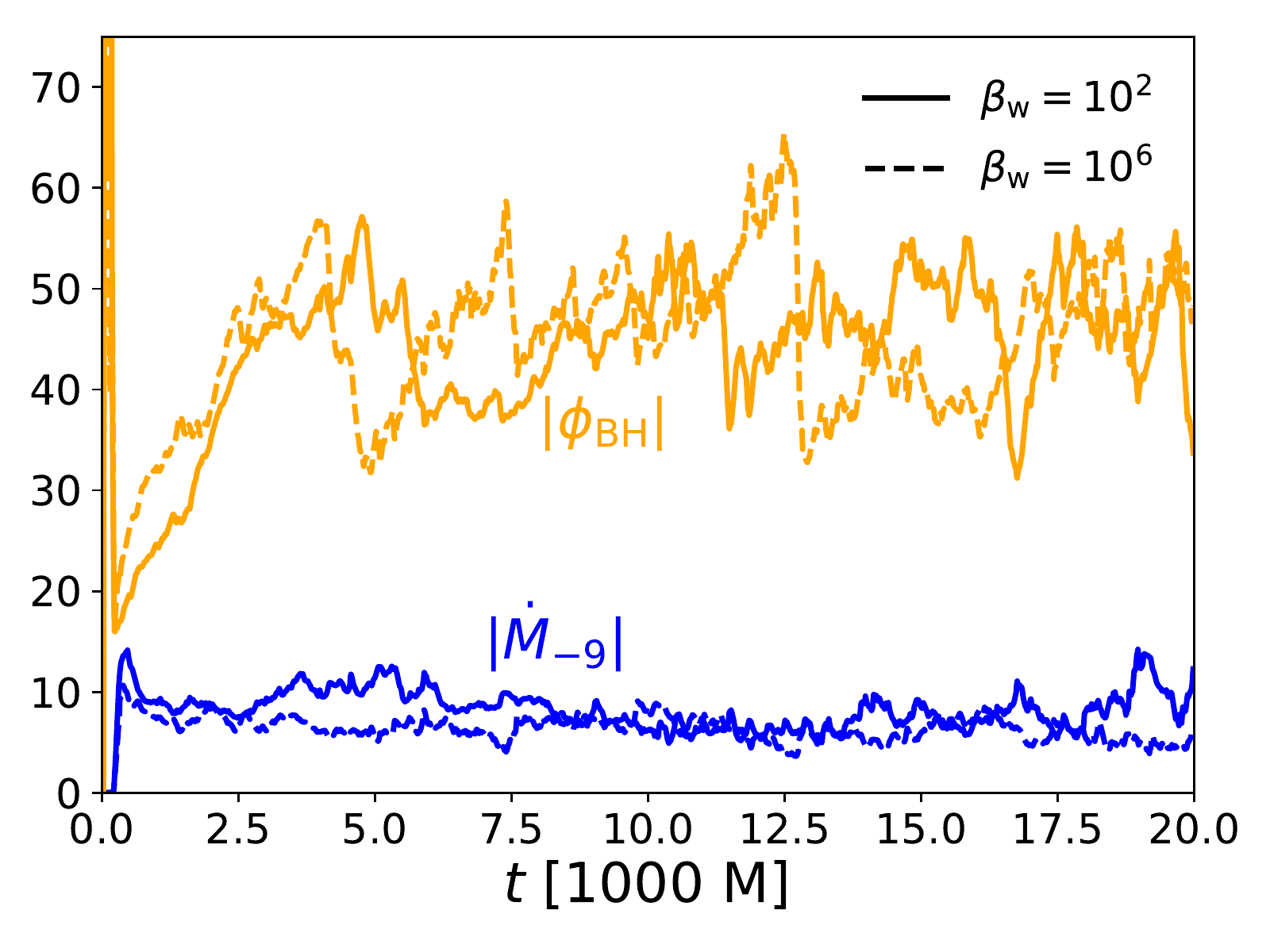}
\caption{Accretion rate at the event horizon in units of $10^{-9}$ $M_\odot$ yr$^{-1}$ (blue), $|\dot M_{-9}|$, and dimensionless flux threading the event horizon (orange), $\phi_{\rm BH}$,  vs. time in our GRMHD simulations with $\beta_{\rm w} = 10^2$ (solid) and $\beta_{\rm w} = 10^6$ (dashed).  The two simulations show  remarkably similar behavior in both quantities despite the $4$ orders of magnitude difference in $\beta_{\rm w}$, the ratio between the ram pressure and the magnetic pressure for the $\sim$ 0.1 pc-scale WR stellar winds in the original R20 simulations. The accretion rate is fairly steady over 20,000 $M$ at a value of $\approx 10^{-8} M_\odot$ yr$^{-1}$, consistent with many previous estimates for Sgr A*.  $\phi_{\rm BH}$ steadily increases during the initial $\sim$ 5,000 $M$ of the simulation but then saturates at approximately the MAD limit of $\sim$ $40$--$60$  at which point the field is strong enough to balance the pressure of the inflowing gas.  
 } 
\label{fig:time_plots}
\end{figure}

Figure \ref{fig:time_plots} plots $\phi_{\rm BH}$ and $\dot M$ as a function of time in our $\beta_{\rm w} = 10^2$ and $\beta_{\rm w} = 10^6$ GRMHD simulations.  The curves for the two simulations show essentially the same behavior, demonstrating that these quantities are robustly determined at small radii independent of $\beta_{\rm w}$.  In contrast to the Newtonian MHD simulations, which never became fully arrested (R20)\footnote{There are several possible reasons why the R20 simulations did not become arrested in contrast to the GRMHD simulations presented here:  1) the inner boundary radius was artificially large compared to the event horizon 2) GR effects were not taken into account, and/or 3) the inner boundary was resolved by only $\sim$ 2 cells in radius, which could potentially enhance the diffusion of magnetic field lines and prevented the arrested state from developing.}, $\phi_{\rm BH}$ grows until the MAD limit of $\approx 40$--$60$ is reached, at which point it oscillates about that range in an arrested state.
Despite this, the net accretion rates are fairly constant around $10^{-8} M_\odot$ yr$^{-1} \approx 10^{-7} \dot M_{\rm Edd}$, where $\dot M_{\rm Edd} = L_{\rm Edd}/(0.1 c^2)$ is the Eddington accretion rate for Sgr A*.   This value falls nicely within the limits derived from polarization measurements \citep{Marrone2007} and in the range of previous estimates that fit models to observations (e.g., \citealt{Shcherbakov2010,Mosci2014,CK2015,Ressler2017,Dexter2020}).  

\begin{figure*}
  \begin{center}
\includegraphics[width=0.45\textwidth]{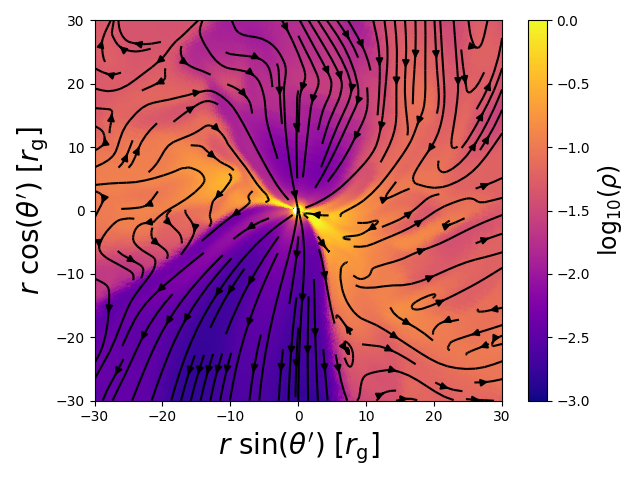}
\includegraphics[width=0.45\textwidth]{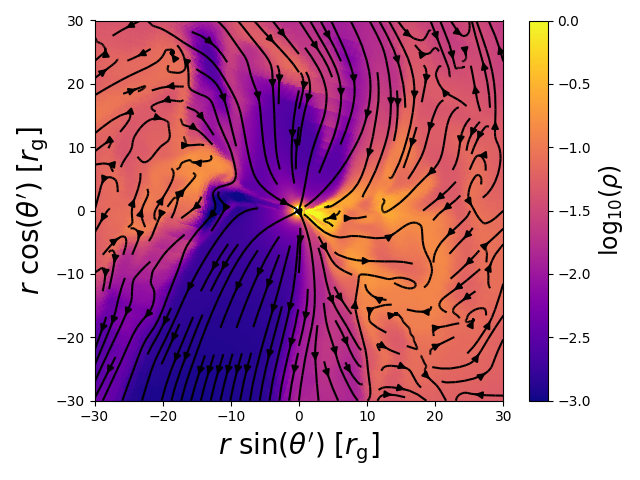}
\end{center}
\caption{2D slices of mass density over-plotted with magnetic field lines in our $\beta_{\rm w} = 10^2$ GRMHD simulation at $t = 10,080 M$ (left) and $t = 11,430 M$ (right).  The frame of the plot is such that the $z^\prime$-axis (i.e., the vertical axis) is aligned with the average angular momentum direction of the gas being fed from large radii ($\sim$ $6 \times 10^{-4}$--$3 \times 10^{-2}$ pc $\approx$ $3 \times 10^3$--$2 \times 10^5$ $r_{\rm g}$) via the WR stellar winds.   A slight misalignment of the magnetic field with this axis causes a $\sim 10$--$20^\circ$ tilt in the density distribution.  The field is strong enough to concentrate the initially quasi-spherical distribution of gas into a disk-like structure.  Furthermore, at several times (e.g., $t = 11,430 M$ in the right panel, see also the peak at this time in $\phi_{\rm BH}$ in Figure \ref{fig:time_plots}), the field strength is sufficiently large to completely push away the accreting gas on one side of the black hole, a defining characteristic of MAD simulations.   Plots from the $\beta_{\rm w}=10^6$ simulation look qualitatively very similar.  Animations: \href{https://smressle.bitbucket.io/animations.html}{https://smressle.bitbucket.io/animations.html}} 
\label{fig:contour_plots}
\end{figure*}

Additional evidence for the magnetically arrested nature of the flow is found in Figure \ref{fig:contour_plots}, which presents 2D contours of mass density over-plotted with magnetic field lines at two different times in our $\beta_{\rm w} = 10^2$ simulation.  In these plots the polar axis coincides with the average angular momentum vector of the gas being fed in from large radii ($\sim$ $6 \times 10^{-4}$--$3 \times 10^{-2}$ pc $\approx$ $3 \times 10^3$--$2 \times 10^5$ $r_{\rm g}$).  The right panel of Figure \ref{fig:contour_plots} shows the gas getting pushed outwards from the left side of the black hole.  This behavior is observed in the simulation sporadically whenever $\phi_{\rm BH}$ reaches a peak (see Figure \ref{fig:time_plots}) and is typical of MAD simulations (e.g., \citealt{Sasha2011,Narayan2012}).  Also consistent with past work (e.g., \citealt{McKinney2012} ),  the gas in Figure \ref{fig:contour_plots} is confined to a relatively thin, turbulent, disk-like structure within $r \lesssim 10 r_{\rm g}$.  This is caused by the strong magnetic field ``choking'' the accretion flow and evacuating the polar regions of matter.   

There are, however,  some key differences between our simulations and previous torus-based MADs.  The non-axisymmetric way in which accretion is fed via spiral-shaped streams (see Figure 11 in R20) leads to one side of the disk being consistently thicker than the other, with the thinner side tending to be, on average, outflowing, at least for $r\gtrsim$10--20$r_{\rm g}$.  In fact, it is always the thinner side of the disk that gets dramatically pushed outward after $\phi_{\rm BH}$ reaches a maximum (e.g., the right panel of Figure \ref{fig:contour_plots}); such events in torus-based MADs generally occur on both sides equally. Furthermore, the disk is tilted with respect to the initial angular momentum axis, with the magnitude of the tilt varying from  $\sim$ $20$--$30^\circ$ to $\sim$ 0$^\circ$ over the course of the simulation.   The tilt is caused by the net magnetic field direction being inclined with respect to the initial rotation axis, so that as the field accretes, it becomes strongest at the magnetic pole and pushes the gas towards the magnetic midplane.  This is not to be confused with a black hole spin related tilt (e.g., \citealt{Fragile2005,Liska2018,White2019b}), a possibility to be explored in future work.  Here $a=0$.  

The $\beta_{\rm w}=10^6$ simulation behaves qualitatively similar to its $\beta_{\rm w}=10^2$ counterpart in Figure \ref{fig:contour_plots}, with the main difference being that the tilt is now $\sim$ $90^\circ$.   Again, this is caused by an initial misalignment of the magnetic field direction with the angular momentum axis, but with a larger magnitude.  Generally, we find that large tilts develop more often when the magnetic field is weaker in the WR winds at large radii.  However, even for the $\beta_{\rm w} = 10^2$ case, near 90$^\circ$ tilts are seen at some times. A more detailed discussion of the magnetic field direction in the R20 simulations can be found in Appendix \ref{app:B_dir}.

\subsection{230 GHz Images and Polarization}

\begin{figure*}
  \begin{center}
\includegraphics[width=0.49\textwidth]{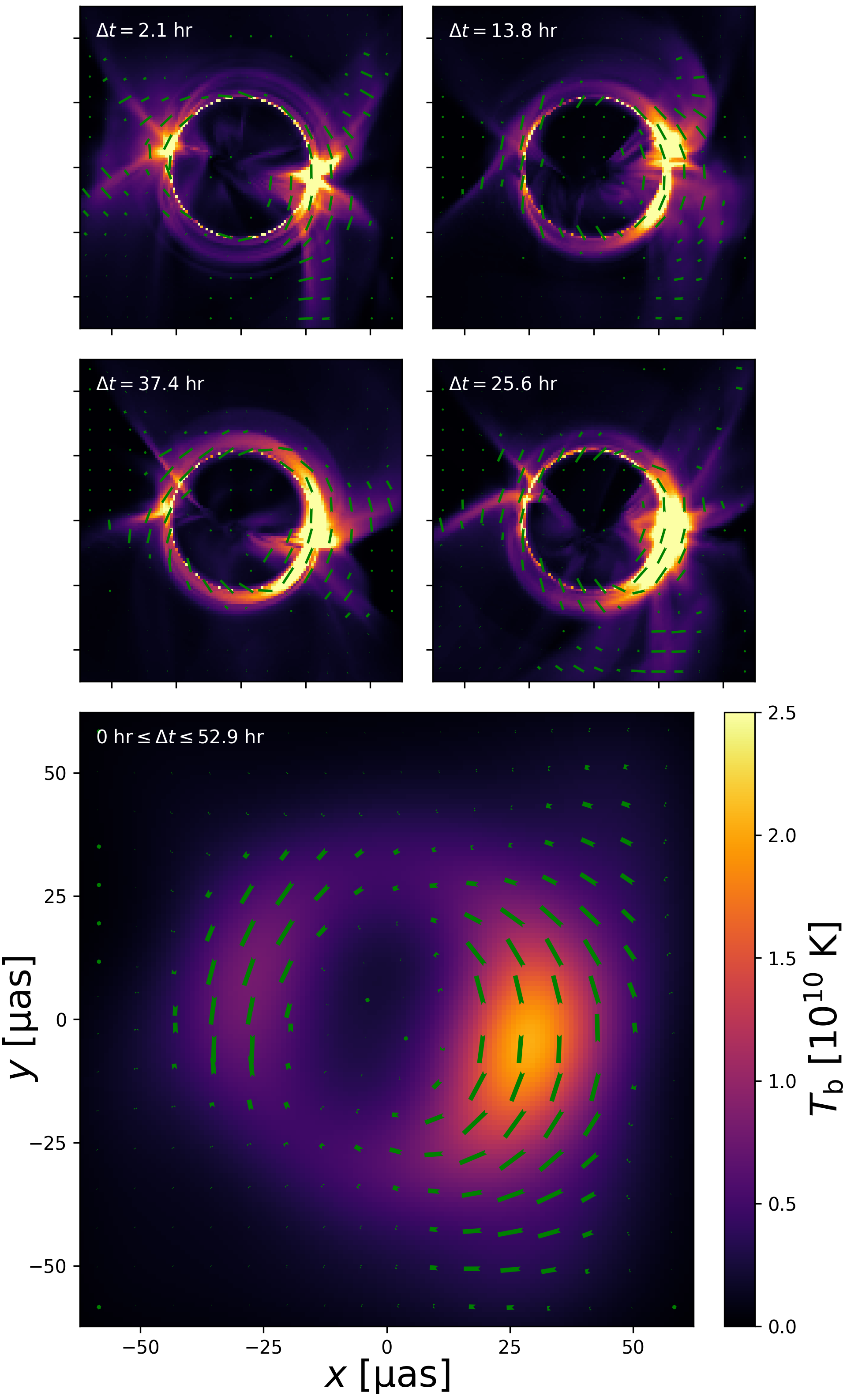}
\includegraphics[width=0.49\textwidth]{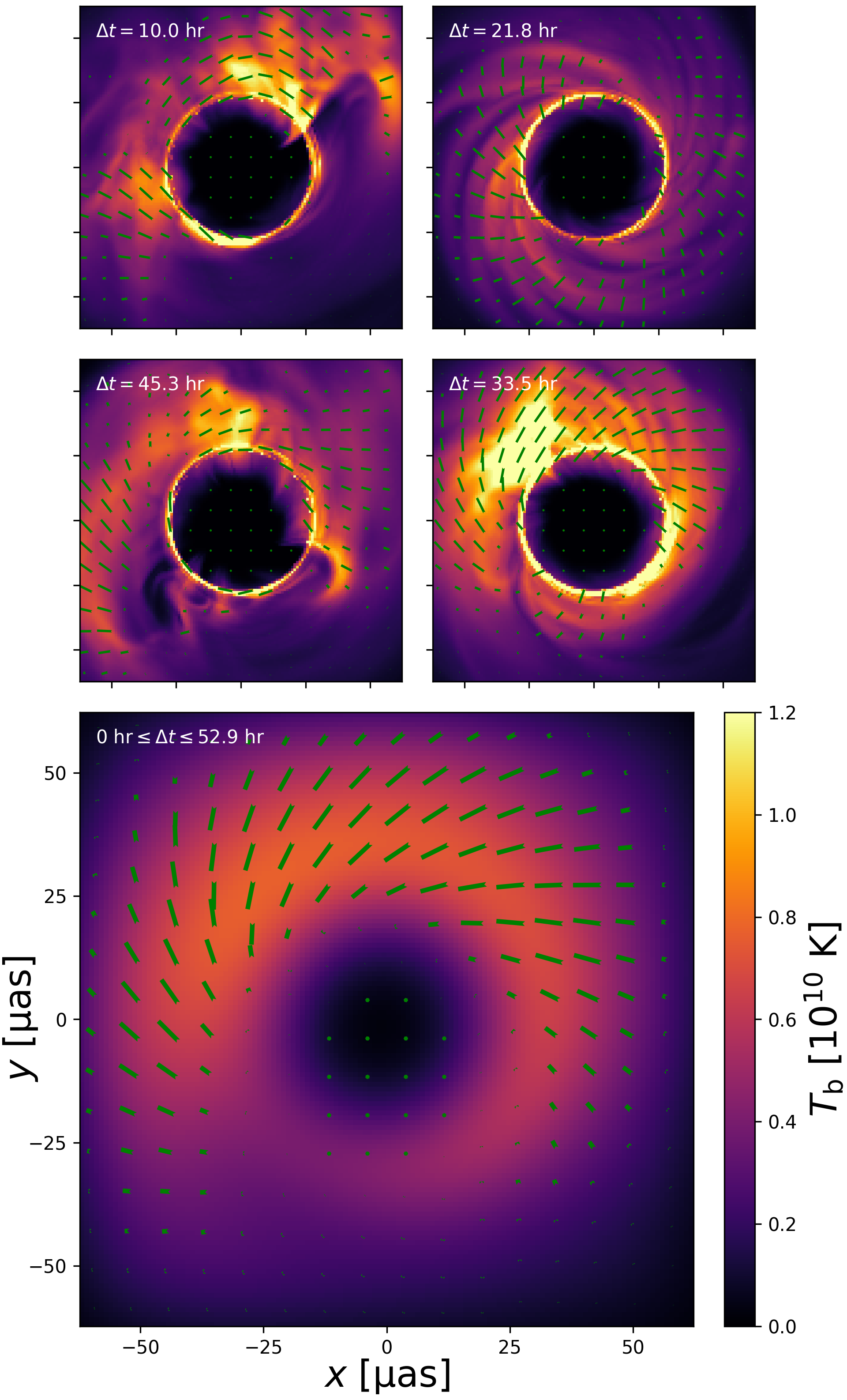}
\end{center}
\caption{230 GHz images and polarization vectors produced from our simulations over the $\sim$ 53 hr interval $11,000$--$20,000 M$ using {\tt grtrans}.  Left: $\beta_{\rm w}=10^2$. Right: $\beta_{\rm w}=10^6$.  The top four images in each column are snapshots proceeding forward in time clockwise starting at the upper left quadrant, labeled by $\Delta t = t - 11,000 M$.  The bottom, larger image in each column is time-averaged and smoothed over a 20 $\mu$as Gaussian.  Polarization vectors are scaled to be proportional to polarization fraction times intensity.  
The $\beta_{\rm w}=10^2$ images are relatively edge-on with emission stronger to the west, while the $\beta_{\rm w}=10^6$ images are relatively face-on with more symmetrically distributed emission.   
These orientations are set by the directions of the net magnetic field being fed from large radii, which is sensitive to the particulars of the R20 wind-fed, larger scale MHD simulations.   
For the chosen electron model (\S \ref{sec:methods}) internal Faraday rotation effects are relatively weak so that polarization vectors are coherent, tracing the underlying magnetic fields.  Animations: \href{https://smressle.bitbucket.io/animations.html}{https://smressle.bitbucket.io/animations.html}} 
\label{fig:image}
\end{figure*}

Figure \ref{fig:image} shows 230 GHz images over-plotted with polarization vectors computed from our two simulations.    Contained in Figure \ref{fig:image} are both time-averaged images and snapshots over the interval $11,000$--$20,000 M$ $\sim$ 53 hr.  The time-averaged image has has been blurred over a 20 $\mu$as Gaussian to mimic the Event Horizon Telescope (EHT) resolution \citep{EHT1}.   For $R_{\rm high} = 46$  ($\beta_{\rm w}=10^2$) and $R_{\rm high} = 10$ ($\beta_{\rm w}=10^6$), the time-averaged fluxes at 230 GHz are 2.4 Jy, consistent with observations \citep{Doeleman2008}.  In contrast to previous work, the orientation of these images with respect to Earth is not a free parameter but is determined by the direction of the net magnetic field being fed from large radii via stellar winds.  Unfortunately, this orientation is sensitive to the precise time used in the R20 simulations as initial conditions, especially for $\beta_{\rm w}=10^6$ (see Appendix \ref{app:B_dir} for a discussion and the right panel of Figure \ref{fig:lb_angle}), and thus is not a robust prediction of our model.   Over the course of the $\sim$ 5 day duration of our simulations, the angular momentum vector of the gas can shift by $\sim$ 10--20$^\circ$, while much larger changes could occur on $\gtrsim 10$ yr time-scales (especially for $\beta_{\rm w} = 10^6$).
For our fiducial $\beta_{\rm w}=10^2$ simulation, we find nearly edge-on inclinations ($i \sim $290--300$^\circ$, where $i$ is the angle that the inner disk makes with the line of sight), tilted by $\sim 20$--$30^\circ$ with respect to the clockwise stellar disk.
As a result, a strong Doppler boost is present on the west side of the images in Figure \ref{fig:image}. 
Conversely, the orientations of the images generated from our fiducial $\beta_{\rm w}=10^6$ simulation are essentially face-on ($i \sim -90^\circ$) and are thus less influenced by Doppler effects. 
This lack of Doppler boosting in the $\beta_{\rm w} =10^6$ simulation combined with the fact that the $\beta_{\rm w}=10^2$ simulation has densities,  (total) temperatures, and magnetic field strengths that are each $\sim$ $50 \%$ higher near the horizon explains why the $R_{\rm high}$ needed to achieve a time-averaged 2.4 Jy flux is 4.6 times smaller for $\beta_{\rm w}=10^6$ compared to $\beta_{\rm w}=10^2$.  Note that the emission-weighted $\langle T_e\rangle$ is comparable for both simulations, $k_{\rm B} T_e / m_e c^2$ $\approx$17 for $\beta_{\rm w} = 10^2$ and $\approx$ 22 for $\beta_{\rm w} = 10^6$, where $k_{\rm B}$ is Boltzmann's constant and $m_e$ is the electron mass. For comparison, the emission-weighted magnetic field strength and density are $\approx$ 20 G and $\approx 3.7 \times 10^5$ cm$^{-3}$ for $\beta_{\rm w}=10^2$ and $\approx $ 10 G  and $\approx$ $2.3 \times 10^{5}$ cm$^{-3}$ for $\beta_{\rm w}=10^6$.
The images from both simulations show interesting time variability over the course of the $\sim$ $53$ hours, with bright spots appearing, disappearing, brightening, dimming, and even orbiting in the case of $\beta_{\rm w}=10^6$.   This highlights one of the challenges for EHT in imaging Sgr A*.  

The polarization vectors in Figure \ref{fig:image} are coherent and ordered for both simulations, tracing out the ordered magnetic field.  Internal Faraday rotation (i.e. Faraday rotation on the scale of the image) is weak enough to prevent depolarization.
Integrated over the entire image, the linear polarization fractions are $6.8 \pm 4.0$\%  ($\beta_{\rm w}=10^2$) and $6.1 \pm 3.4$\% ($\beta_{\rm w}=10^6$) across the $\sim$ 53 hour time window.  These values are in excellent agreement with the mean values of 3.6--7.8\%  reported by \citet{Bower2018}.  We also find the emission to have a small degree of circular polarization (CP) provided mainly $(\gtrsim 90 \%)$ by  Faraday conversion of initially linearly polarized light.  The CP fractions are $0.18 \pm 0.15$\%  ($\beta_{\rm w}=10^2$) and $0.35 \pm 0.15$\% ($\beta_{\rm w}=10^6$),  low compared to the $1.2 \pm 0.3 \%$ reported by \citet{Munoz2012}.  We note that the simulated polarization is somewhat sensitive to both the electron temperature model and assumed abundance ratios.

\section{Discussion And Conclusions}
\label{sec:disc_conc}
We have presented the results of 3D GRMHD simulations of Sgr A* in which the properties of the flow near the horizon are directly linked to the WR stellar winds feeding the Galactic center on $\gtrsim$ $10^5 r_{\rm g}$ scales.  
This was done by refining at small radii in our previously published MHD simulations on $10^{-3}$--1 pc scales in which the WR stars are treated as source terms in mass, momentum, energy, and magnetic field (R20, see Appendix \ref{app:inits}).  Since the properties of the winds are chosen to conform to all known observational constraints, our results are much more predictive than previous GRMHD simulations which start with somewhat ad hoc initial conditions.  Case in point, we have only a limited number of free parameters, namely: the spin of the black hole, $a$; the ratio between the ram pressure and the magnetic pressure in the WR stellar winds, $\beta_{\rm w}$; and the orientation of the spin axes of the WR stars (which determine the orientation of the field in the winds).  Many (but not all) of our results are insensitive to $\beta_{\rm w}$ because the magnetic field tends to reach equipartition with the thermal pressure by the time it reaches the horizon (confirming the extrapolation done by R20) so that its strength at horizon scales is independent of the initial value.  Our results are also mostly insensitive to the choice of spin axes because the orientation of the field in the inflowing gas is primarily determined by the orbital angular momentum vector of the star from which it was emitted, not the initial orientation of the field with respect to the star (Appendix \ref{app:B_dir}).  In calculating emission we have an additional degree of freedom in choosing an electron temperature model.

These simulations smoothly connect with the R20 simulations (Appendix \ref{app:3_sim_results}) that were consistent with the observed X-ray luminosity at $2$--$10^{\prime\prime}$ scales \citep{Baganoff2003}, rotation measure towards Sgr A* \citep{Marrone2007,Bower2018}, and inferred density power law index ($\rho$ $\tilde \propto$ $r^{-1}$, \citealt{Gillessen2019}). In the current work,  we find that the accretion rate through the horizon, $\sim 10^{-8} M_\odot$ yr$^{-1}$ (Figure \ref{fig:time_plots}),  is also consistent with observational estimates \citep{Marrone2007}.  We stress that our prediction of $\dot M$ has essentially no free parameters, so this consistency with observations is very non-trivial.  The combination of this work and R20 thus forms a self-consistent solution for the accretion flow that agrees well with several independent observational probes across many orders of magnitude in radius.

We find that the horizon scale accretion flow becomes magnetically arrested, with the dimensionless flux threading the black hole saturating at the MAD limit of $\sim$ 40--60 (Figure \ref{fig:time_plots}).
This strong concentration of magnetic field restricts the accretion flow to a relatively thin, disk-like configuration that, unlike traditional torus-based MADs in the literature, is tilted with respect to the initial angular momentum axis of the gas (Figure \ref{fig:contour_plots}).  The magnitude of the tilt corresponds to the angle that the net field makes with the rotation axis on larger scales in the original MHD wind-fed simulations.  This tilt is more likely to be large ($\gtrsim 45^\circ$) for more weakly magnetized WR stellar winds; but even at a particular $\beta_{\rm w}$ its value varies in time.  At times the radial Lorentz force provided by the field is even strong enough to completely expel sections of the disk, behavior typical of MAD simulations.   
A MAD in the Galactic center has been suggested as one possible explanation for the recent GRAVITY  observations that show temporal variability in the polarization vector of near infrared flares consistent with poloidal magnetic fields \citep{GRAVITYFlare}.   

The horizon-scale angular momentum of the gas in our simulations is set by the net magnetic field direction of the material being fed in from large radii.  Because of this, even though the angular momentum direction at large radii is rarely different from that of the stellar disk (Figure 9 in R20), the orientation of the flow at small radii can be in an entirely different direction.
Case in point, our fiducial $\beta_{\rm w}=10^2$ simulation is close to edge-on with an inclination angle of $\sim$ 290--300$^\circ$ while our fiducial $\beta_{\rm w} = 10^6$ simulation is nearly face-on with an inclination angle of $\sim$ $180^\circ$.  The contrast is evident in the corresponding 230 GHz images (Figure \ref{fig:image}).   The $\beta_{\rm w}=10^2$ images display a clear asymmetry with emission focussed on the west caused by Doppler boosting while the $\beta_{\rm w}=10^6$ images are more symmetric. Note, however, that $\beta_{\rm w}$ alone is not enough to determine whether the horizon-scale gas in our simulations ends up in a configuration closer to edge-on or closer to face-on, as both $\beta_{\rm w}=10^2$ and $\beta_{\rm w}=10^6$ can be either depending on the particular time in the R20 simulations used to generate the GRMHD initial conditions (see Appendix \ref{app:B_dir} for a discussion). 
As a result, we cannot robustly predict the orientation of the image ``today'' in Sgr A*. 
Images from both simulations can also look significantly different depending on when they are observed, even over the course of a day (see Figure \ref{fig:image}).   
Our simulations have relatively weak internal Faraday rotation so that the polarization vectors (Figure \ref{fig:image}) are well ordered, tracing out the magnetic field structure.    The linear polarization fraction is time variable and depends on the post-processing electron temperature model but is, on average, consistent with measurements of Sgr A*.  

We have limited this initial study to non-spinning black holes ($a=0$).  If Sgr A* is rapidly rotating, several of the properties of our simulations could change.  
It is likely that a strong \citet{BZ1977} jet would develop and potentially alter the accretion rate or flow structure.  This is especially true given the magnetically arrested nature of the flow; future simulations could help constrain the magnitude of $a$ given the lack of direct evidence for a radio jet in Sgr A*.  Moreover, if the rotation axis of the black hole is misaligned with the magnetic polar axis, the innermost gas and magnetic field could be warped and perhaps even align with the spin of the black hole, altering the structure of the images \citep{Liska2018,White2019b,White2020,Chatterjee2020}.   We will explore $a \ne 0$ in future work.
 
\section*{Acknowledgments}
We thank the anonymous referee for a thoughtful and detailed report. We thank O. Blaes, J. Dexter, and C. Gammie for useful discussions, as well as all the members of the Horizon Collaboration, \href{http://horizon.astro.illinois.edu}{http://horizon.astro.illinois.edu}. SMR was supported by the Gordon and Betty Moore Foundation through Grant GBMF7392.  SMR also thanks R. and D. Ressler for
their generous hospitality during part of the writing of this manuscript.
This work was supported in part by NSF grants NSF PHY--1748958,  AST--1715054, AST--1715277, a Simons Investigator award from the Simons Foundation, and by the NSF
through XSEDE computational time allocations TG--AST090038
on SDSC Comet and TG--AST170012 on Stampede2.  This work was made possible by computing time granted by UCB on the Savio cluster.

\bibliographystyle{aasjournal}
\bibliography{mad}

\appendix
\section{Generating Initial Conditions For GRMHD From R20 wind-fed Accretion Simulations}
\label{app:inits}

In this Appendix we describe our method for creating initial/boundary conditions for a GRMHD simulation from one of the larger scale, wind-fed simulations of R20 using intermediate scale MHD simulations.

These intermediate simulations extend from the event horizon out to a radius just inside the orbits of the WR stars closest to the black hole.  More precisely, they encompass a (0.0625 pc)$^3$ cube in Cartesian coordinates centered on the black hole, covered by a 128$^3$ base resolution and 11 additional levels of nested SMR designed to mimic logarithmic spacing in radius.  Approximately every factor of 2 decrease in radius the grid spacing halves, so that the highest level of refinement is $\Delta x _{\rm min} \approx 2.4 \times 10^{-7} $ pc $\approx$ 1.1 $r_{\rm g}$. This domain overlaps with the domain of the wind-fed accretion simulations described in R20, which extend from about $300 r_{\rm g}$ out to a radius just outside the orbits of most of the WR stars ($\sim 1$ pc).  
The overlapping domains allow us to use the results of the R20 simulations as initial conditions for this new smaller scale simulations.  In particular, we focus mainly on the $\beta_{\rm w}=10^2$ simulation, where $\beta_{\rm w}$ is the ratio between the ram pressure and the magnetic pressure in each wind, but also use data from the $\beta_{\rm w} = 10^6$ simulation for comparison.  We discuss the effect of varying $\beta_{\rm w}$ in \S \ref{sec:disc_conc} of the main text.  Data is taken at $t=0.15$ kyr, that is, 0.15 kyr from the present day for $\beta_{\rm w} = 10^2$ and $t = 0.05$ kyr (0.05 kyr from the present day) for $\beta_{\rm w} = 10^6$. These times were chosen because they fall within intervals of the simulations during which the angular momentum of the gas is roughly constant in time and radius (see Figure 9 in R20), aligned with the clockwise stellar disk.   Though precise details of the analysis we present will depend on this choice, we believe that the results should be representative of all times and that the conclusions we draw are robust.

Interpolation onto the new grid is done using the nearest neighbor method for the hydrodynamic variables $\rho$ (mass density), $ P$ (pressure), and $\mathbf{v}$ (velocity), while $\mathbf{B}$ (magnetic field) is initialized from the vector potential  $\mathbf{A}$ via $\mathbf{B} = \mathbf{\nabla} \times \mathbf{A}$, where $\mathbf{A}$ is obtained by solving the vector Poisson equation $\mathbf{\nabla}^2 \mathbf{A} = - \mathbf{\nabla} \times \mathbf{B}$ on the original grid and then interpolating onto the new grid.  To minimize any artificial effects of the original inner boundary, we use simulation data only from $r \ge 10^{-3}$ pc, with cells $r <10^{-3}$ pc being set to the numerical floors in density/pressure, zero velocity, and zero magnetic field.\footnote{Technically, the magnetic field is set from the original vector potential weighted by an exponentially decreasing function of decreasing radius which rapidly approaches zero below $r  = 10^{-3}$ pc.}  
 Furthermore, we define an effective inner boundary for the intermediate simulations as the cells within $r_{\rm in}$ = $2  \Delta x_{\rm min} \approx 5 \times 10^{-7} $ pc $\approx$ 2.2 $r_{\rm g}$; within $r_{\rm in}$ all cells are set to the numerical floors in density/pressure and zero velocity while the magnetic field is allowed to freely evolve (that is, given the floored fluid variables the induction equation is solved without any modification).  In the past, we experimented with more sophisticated treatments of the inner boundary, such as spherical inflow-like conditions or radial extrapolation. These methods, however, showed no significant improvements on test problems (e.g., spherical Bondi inflow) nor did they effect the qualitative nature of our wind-fed accretion simulations.  The outer (cubic) boundary of the grid is fixed to the initial conditions and does not change with time. 
  
For these MHD simulations, Newtonian, point source gravity is included for a black hole of mass $M = 4.3 \times 10^{6} M_\odot$ \citep{Gillessen2017}.\footnote{For consistency with previous simulations we use this value for the mass of Sgr A* instead of the updated estimate based on the pericenter passage of S2 \citep{GravityS2}.  } 
 The simulations are run for 0.24 yr $\sim 3.4 \times 10^5 r_{\rm g}$, or approximately 1.5 free-fall times at $r = 10^{-3}$ pc.   Since this is much shorter than the $\sim$ 25 yr free-fall time at the outer boundary, the assumption of static outer boundary conditions is justified. The adiabatic index of the gas is $\gamma=5/3$.  Radiative cooling is inefficient for the radii encompassed by the simulations and is not included.

The combination of the original, larger scale, wind-fed accretion simulations with these smaller scale, re-initialized simulations essentially provides us with a self-consistent MHD accretion model over the entire radial range of interest, albeit without the inclusion of general relativistic effects and with the innermost $\sim$ 35$r_{\rm g}$ relatively unresolved.   The re-initialized MHD simulations then serve as the initial and boundary conditions for GRMHD simulations in an analogous way to how the R20 simulations served as initial and boundary conditions for the re-initialized MHD simulations.   We interpret the MHD $\rho$, $P$, and $\mathbf{v}$ as the GRMHD rest frame density, pressure, and the spatial components of the four velocity, $u^i$, respectively, and interpolate these onto the Cartesian GRMHD grid described in \S \ref{sec:methods} of the main text.  We again solve the vector Poisson equation for $\mathbf{A}$ and interpret it as $A_i$, which is interpolated onto the new grid and used to generate the magnetic field via $B^i = \epsilon^{ijk}\partial_j A_k$, where $\epsilon^{ijk}$ is the Levi--Civita tensor.  These initial conditions are used only for $r\ge 50 r_{\rm g}$ where relativistic effects are small; for $r < 50 r_{\rm g}$ the density and pressure are initialized to the numerical floors, the four-velocity is free-fall, and the magnetic field is zero.  

Our GRMHD simulations are performed in Cartesian Kerr--Schild (CKS, \citealt{Kerr1963}) coordinates using the user-defined coordinate module in {\tt Athena++}. 
In terms of the Kerr--Schild $r,\theta$, and $\varphi$, these are \citep{Kerr1963}\footnote{Note that in the original paper by Kerr there was an error in the sign of $a$ \citep{Kerr2007} so that, in his expressions, $a>0$ describes a black hole with angular momentum pointing in the $-z$ direction.  We have altered our expressions so that $a>0$ corresponds to a black hole with angular momentum pointing in the $+z$ direction.  } 
\begin{eqnarray}
  x = r \sin(\theta) \cos(\varphi) + a\sin(\theta)\sin(\varphi) \\
  y = r \sin(\theta) \sin(\varphi) - a \sin(\theta)\cos(\varphi) \\
  z = r \cos(\theta),
\end{eqnarray}
where $a$ is the spin of the black hole.  The metric and inverse metric in CKS coordinates are
\begin{eqnarray} 
  g_{\mu \nu} = \eta_{\mu \nu} + f l_\mu l_\nu \\
  g^{\mu\nu} = \eta^{\mu \nu} - f l^\mu l^\nu,
\end{eqnarray}
where $\eta_{\mu \nu}$ is the Minkowski metric and 
\begin{eqnarray}
f = \frac{2r^3}{r^4 + a^2 z^2}\\
l_\mu = \left(1, \frac{rx+ay}{r^2+a^2}, \frac{ry-ax}{r^2+a^2},\frac{z}{r}\right) \\
l^\mu = \eta^{\mu\nu} l_\nu .
\end{eqnarray}   
Derivatives of the metric are computed analytically to calculate the connection coefficients.

This technique is outlined schematically in Figure \ref{fig:domain}, which shows the radial extent of all three simulations and the range of simulation data used to initialize the MHD and GRMHD simulations.   The location of the WR stars is also indicated for reference.

\section{Effectiveness of the Three-Simulation Technique}
\label{app:3_sim_results}
In this Appendix we demonstrate that the three-simulation technique described in Appendix \ref{app:inits} produces a consistent solution across the $\gtrsim$ 6 orders of magnitude in radius.

\begin{figure}
\includegraphics[width=0.45\textwidth]{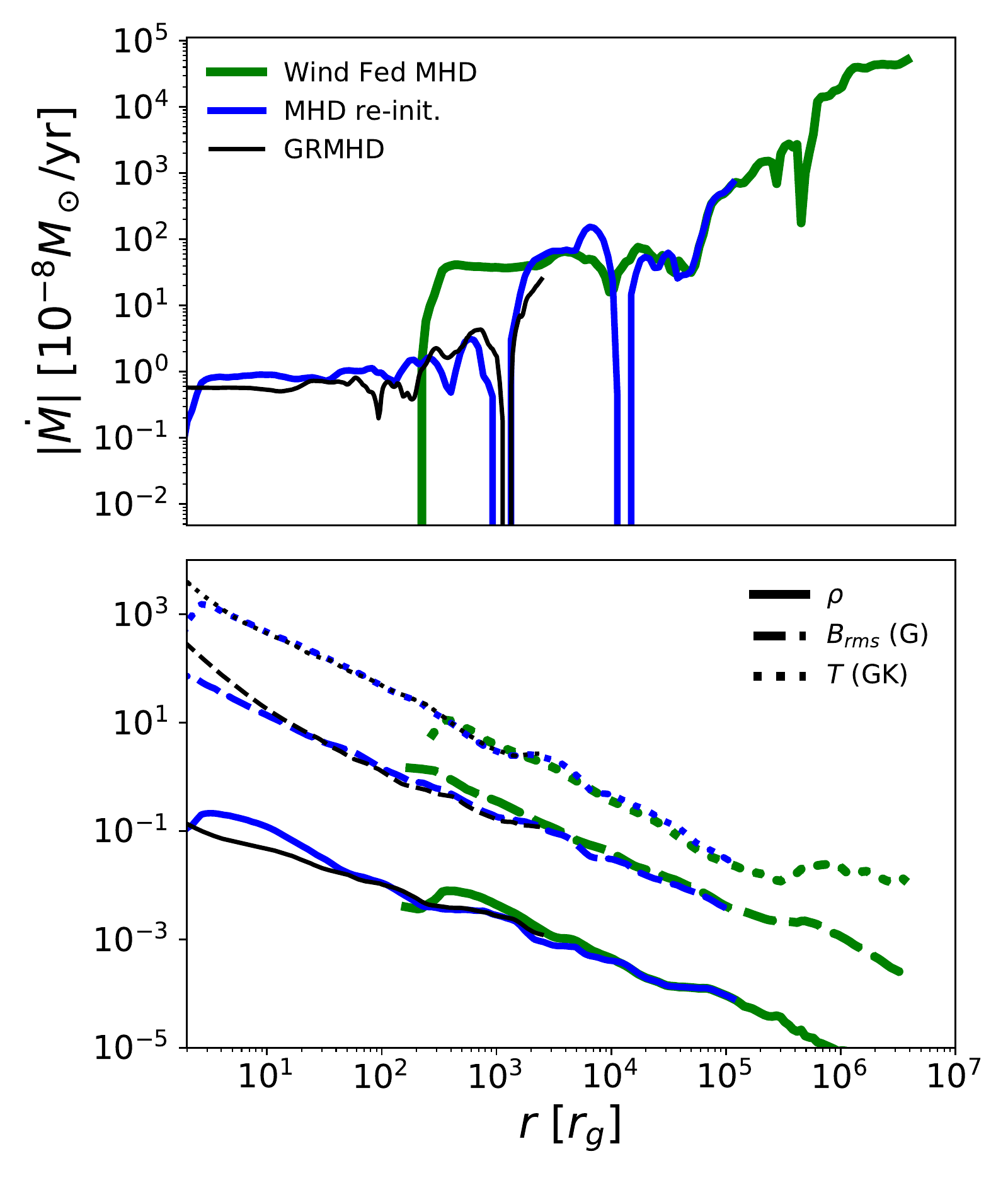}
\caption{Radial profiles of angle-averaged quantities in our three different $\beta_{\rm w}=10^2$ simulations (see Appendix \ref{app:inits} and Figure \ref{fig:domain}).  The R20 wind-fed MHD simulation is green (with data plotted at $t = 0.15$ kyr), the MHD simulation re-initialized from R20 data is blue (with data plotted at $t = 0.15$ kyr + $0.24$ yr), and the GRMHD simulation is black (with data plotted at $t = 0.15$ kyr + $0.24$ yr + $74$ hr, where $74$ hr $\approx$ 12,000 $M$).  Top: Accretion rate, $|\dot M|$, in units of $10^{-8}$ solar masses per year.  Bottom: mass density, $\rho$, temperature in $10^9 K$, $T$, and root-mean-squared magnetic field strength in Gauss, $B_{\rm rms} \equiv \sqrt{\langle b^2 \rangle}$.  The radial profiles of $\rho$, $T$, and $B_{\rm rms}$ all follow power laws consistent across all three simulations.  On the other hand, the accretion rate in the re-initialized MHD simulation and the GRMHD simulation is reduced by almost 2 order of magnitude from the larger scale, wind-fed MHD simulation because of the significantly reduced size of the inner boundary which reduces the maximum angular momentum able to accrete.   The reduction in accretion rate is qualitatively consistent with the extrapolation presented in R20 and, together with the radial profiles, shows that our method of re-initializing simulations at smaller scales is behaving self-consistently and as expected.  } 
\label{fig:rad_profiles}
\end{figure}

Figure \ref{fig:rad_profiles} shows the angle-averaged radial profiles of accretion rate, $\dot M$, mass density, $\rho$, temperature, $T$, and magnetic field strength for the three simulations we use to model Sgr A*, including the R20 wind-fed MHD simulation, the re-initialized MHD simulation used to bridge the gap between large and small scales, and the GRMHD simulation, all for $\beta_{\rm w} = 10^2$.  In MHD, we calculate $\dot M$ using $\dot M_{\rm MHD} = -\iint \rho v_r r^2 \sin(\theta) d\theta d\varphi$, where $v_r$ is the radial velocity and $r,\theta,\varphi$ are the standard flat-space spherical coordinates, while for GRMHD we use $\dot M =  -\iint \rho u^r \sqrt{-g_{\rm KS}} \textrm{ }d\theta d\varphi,$ where $g_{\rm KS}$ and $r,\theta,\varphi$ are the determinant of the metric and the coordinates of spherical Kerr--Schild. The solution for the magnetic and hydrodynamic quantities across the three simulations spanning $\sim$ 6--7 orders of magnitude in radius is generally continuous, with $\langle \rho \rangle$, $\langle B_{\rm rms}\rangle$, and $\langle T\rangle_\rho$ all being well approximated by power laws $\tilde\propto$ $r^{-1}$.  The root-mean-squared magnetic field strength, $B_{\rm rms}$, is computed as $\sqrt{\langle |\mathbf{B}|^2\rangle}$ in MHD and $\sqrt{\langle b^\mu b_\mu\rangle}$ in GRMHD, where $b^\mu$ is the magnetic four vector (e.g., \citealt{Gammie2003}). Here $\langle \rangle$ represents volume-weighted angle averages and the subscript $\langle\rangle_\rho$ indicates that the average is weighted by $\rho$.   The specific angular momentum of the gas (not shown) is similarly well behaved to the quantities in Figure \ref{fig:rad_profiles}, with $\langle l \rangle_\rho$ $\approx$ $0.5 l_{\rm kep} \propto \sqrt{r}$ for radii $\lessapprox 0.1$ pc ($\approx 5 \times 10^5 r_{\rm g} $).   

\begin{figure}
\includegraphics[width=0.45\textwidth]{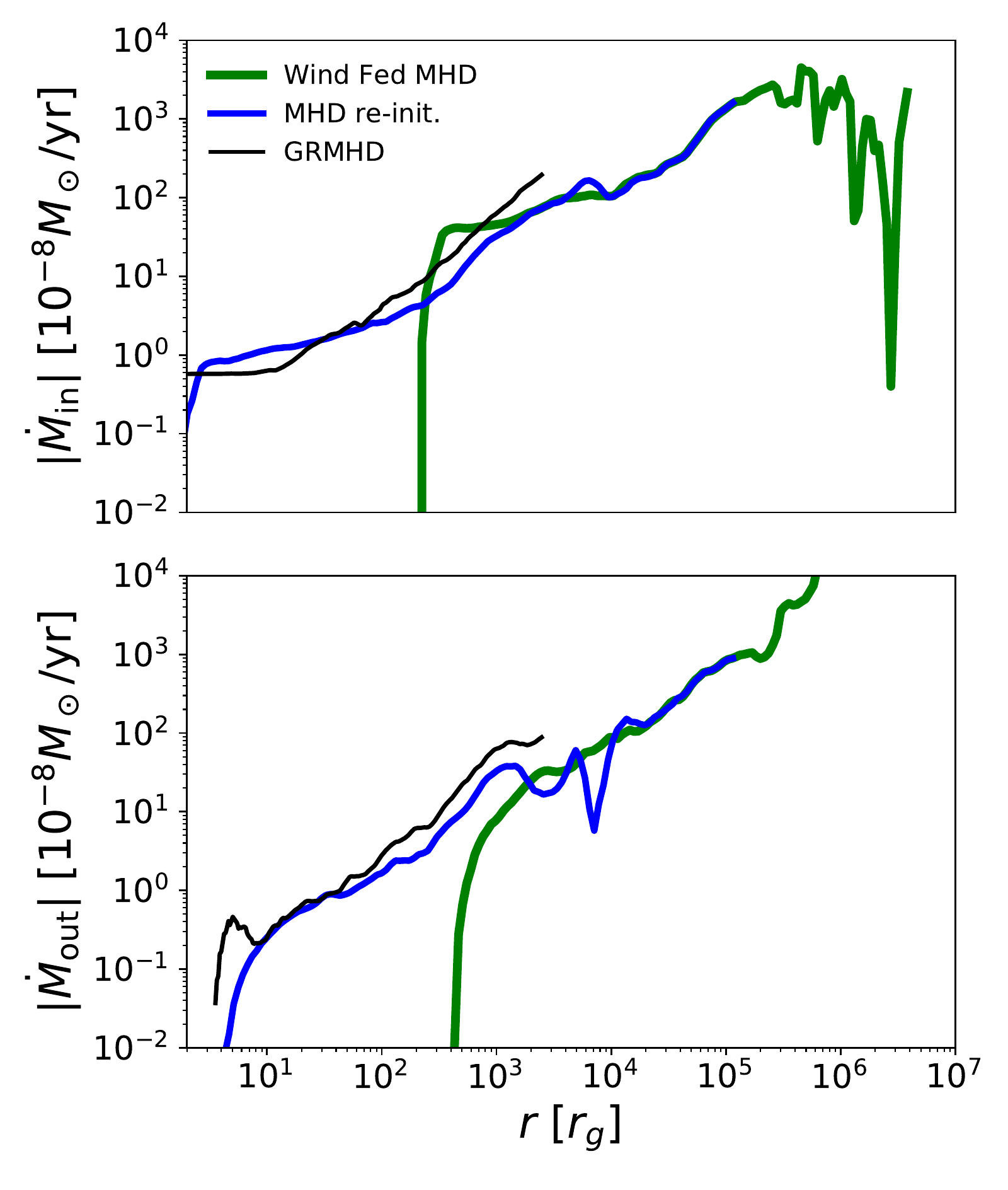}
\caption{Mass inflow (top) and outflow (bottom) rates ($\dot M_{\rm in}$ and $\dot M_{\rm out}$, respectively)  from our three $\beta_{\rm w}=10^2$ simulations as a function of radius. The R20 wind-fed MHD simulation is green (with data plotted at $t = 0.15$ kyr), the MHD simulation re-initialized from R20 data is blue (with data plotted at $t = 0.15$ kyr + $ 0.24$ yr), and the GRMHD simulation is black (with data plotted at $t = 0.15$ kyr + $ 0.24$ yr + $74$ hr, where $74$ hr $\approx$ 12,000 $M$).   Like the density, magnetic field strength, and temperature shown in the bottom panel of Figure \ref{fig:rad_profiles}, both the inflow and outflow rates form essentially continuous power laws across the three simulations.   Throughout most of the domain, $|\dot M_{\rm in}| \approx |\dot M_{\rm out}|$, with the relatively small difference between these two quantities accounting for the net accretion rate, $\dot M$, shown in the top panel of Figure \ref{fig:rad_profiles}.    
 } 
\label{fig:mdot_in_out}
\end{figure}

On the other hand, the average radial velocity, $\langle u^r \rangle_\rho$ (and thus the accretion rate shown in the top panel of Figure \ref{fig:rad_profiles} since $\langle u^r \rangle_\rho \propto |\dot M|r^{-2}$) does not form a continuous power law across the three simulations.\footnote{In the outer radial range of the two smaller scale simulations ($r \gtrsim 10^3$ $r_{\rm g}$), $\dot M$ and $u^r$ agree with the corresponding values of the R20 simulation only because they have not been run long enough for these radii to reach the new equilibrium. }  This is because the effect of the inner boundary (the event horizon in the GRMHD simulation) tends to force the radial velocity to be comparable to the free fall speed at the inner boundary radius, whereas at all other radii the average radial velocity tends to be $\ll$ free fall. In the MHD simulations, this is achieved by the boundary condition limiting outflow and modestly enhancing inflow, while most of the domain is characterized by a balance of inflow and outflow with $\sqrt{\langle v_r^2\rangle} \gg |\langle v_r\rangle|$.   As the boundary radius is decreased to the appropriate value for the event horizon, the region with outflow balancing inflow extends to smaller radii and the accretion rate decreases such that $\dot M$ $\tilde \propto$ $\sqrt{r_{\rm in}}$, where $r_{\rm in}$ is the inner boundary radius.  This follows from $\langle \rho \rangle$ $\tilde \propto$ $r^{-1}$ and $\langle v_r \rangle_\rho (r= r_{\rm in})$ $\tilde \propto$ $v_{\rm ff}(r=r_{\rm in}) \propto 1/\sqrt{r_{\rm in}}$, where $v_{\rm ff}$ is the free-fall speed (see Appendix A in R18 for an analytic derivation).  The net result is that the accretion rate through the event horizon in the GRMHD simulation is reduced from the original wind-fed MHD simulation by almost 2 orders of magnitude.  While this may seem like a dramatic change in the solution, in fact the local $|v_r|$ in MHD is relatively insensitive to the size of the inner boundary and the inflow/outflow rates are roughly continuous power laws across the three simulations.  This is demonstrated explicitly in Figure \ref{fig:mdot_in_out}, which plots $\dot M_{\rm in} = -\iint \rho v_r r^2 \sin(\theta) (v_r <0)d\theta d\varphi$ and $\dot M_{\rm out} = \iint \rho v_r r^2 \sin(\theta) (v_r >0)d\theta d\varphi $ as a function of radius in the three simulations (with the analogous relativistic expressions used for GRMHD).   Both the inflow and outflow rates can be well represented by approximate power laws across the radial range of interest, with the biggest deviation occurring near the inner boundary of the R20 simulation ($\sim 5 \times 10^{2} \lesssim r \lesssim 3 \times 10^3 r_{\rm g}$).  In this region the inflow rate is larger than one would expect from an extrapolated power law while the outflow rate is slightly smaller than one would expect. This is caused by the ``absorbing'' inner boundary removing all pressure support at $r_{\rm in}$, an artificial effect because $r_{\rm in}$ is artificially large.  Once the re-initialized, smaller-scale MHD simulation reaches a rough steady state, however, the $\sim 5 \times 10^{2} \lesssim r \lesssim 3 \times 10^3 r_{\rm g}$ region ``forgets'' the artificial effects of the original inner boundary and the inflow and outflow rates at these radii  become consistent with what one would extrapolate from $r\gtrsim 3 \times 10^3 r_{\rm g}$.   In other words, in the intervening region between the two MHD simulations, the smaller-scale simulation behaves as we would expect the original R20 simulation to behave if the inner boundary radius were significantly reduced. Such was our goal.    Similar behavior is seen in the intervening regions between the smaller-scale MHD simulation and the GRMHD simulation, though to a much lesser extent because $r_{\rm in}$ in the MHD simulation is comparable to the event horizon radius in the GRMHD simulation. 

If both the inflow and outflow rates are thus well behaved across the three simulations (Figure \ref{fig:mdot_in_out}), why then is there such a large discontinuity in the difference between these two quantities (i.e., $\dot M$, the net accretion rate) going from the R20 simulation to the smaller-scale MHD and GRMHD simulations (top panel of Figure \ref{fig:rad_profiles})? This can be understood by considering the nature of the accretion flow, i.e., an inflow/outflow solution in which the individual inflow/outflow rates are approximately equal and individually decrease in magnitude with decreasing radius.  The net accretion rate is determined from these via the size of the inner boundary (i.e., $\dot M \approx \dot M_{\rm in}(r = r_{\rm in})$), meaning that the smaller the inner boundary radius, the smaller the net accretion rate. This is consistent with many other accretion simulations in which inflow roughly balances outflow (e.g., \citealt{Stone1999,Inayoshi2018}).  
Since the inflow/outflow rates are consistent across all simulations, our predicted horizon-scale accretion rate is robust to the particular choices for inner and outer boundaries of the MHD simulations (and thus not dependent on the net $\dot M$ through the inner boundaries of the two MHD simulations ). In fact, based off of the power-law slope of $\dot M_{\rm in}$ (top panel Figure \ref{fig:mdot_in_out}), the $\sim$ 2 orders of magnitude difference in $\dot M$ seen in the R20 simulation compared to the smaller scale MHD and GRMHD simulations is expected.  

\section{Magnetic Field Direction In The R20 MHD wind-fed Simulations}
\label{app:B_dir}

In this Appendix we describe the behavior of the net magnetic field direction in the MHD, wind-fed simulations of R20.  This is important because the resulting orientation of the gas at event horizon scales in the GRMHD simulations described in the main text is set by this direction.  

In R20, the winds of the WR stars are the only source of magnetic field, with the strength in each wind being parameterized by $\beta_{\rm w}$ and the geometry of  the field lines in each wind being determined as follows. Since the stars are orbiting at $\sim$ 0.1--1 $pc$ scales $\gg$ their stellar radii, flux freezing mandates that the field provided by an individual wind is purely in the $\hat \varphi^\prime$ direction, where $\varphi^\prime$ is defined with respect to the rotation axis of the star.  In practice, since we do not know this axis for any of the WR stars, each was chosen randomly at the beginning of the simulations.  Since $\sim $ 1--3 of the winds typically dominate the accretion budget \citep{Cuadra2008} and only a small fraction of each of these winds actually falls towards the black hole (Appendix A in R18), the supply of coherent magnetic flux is relatively large.

\begin{figure*}
\includegraphics[width=0.95\textwidth]{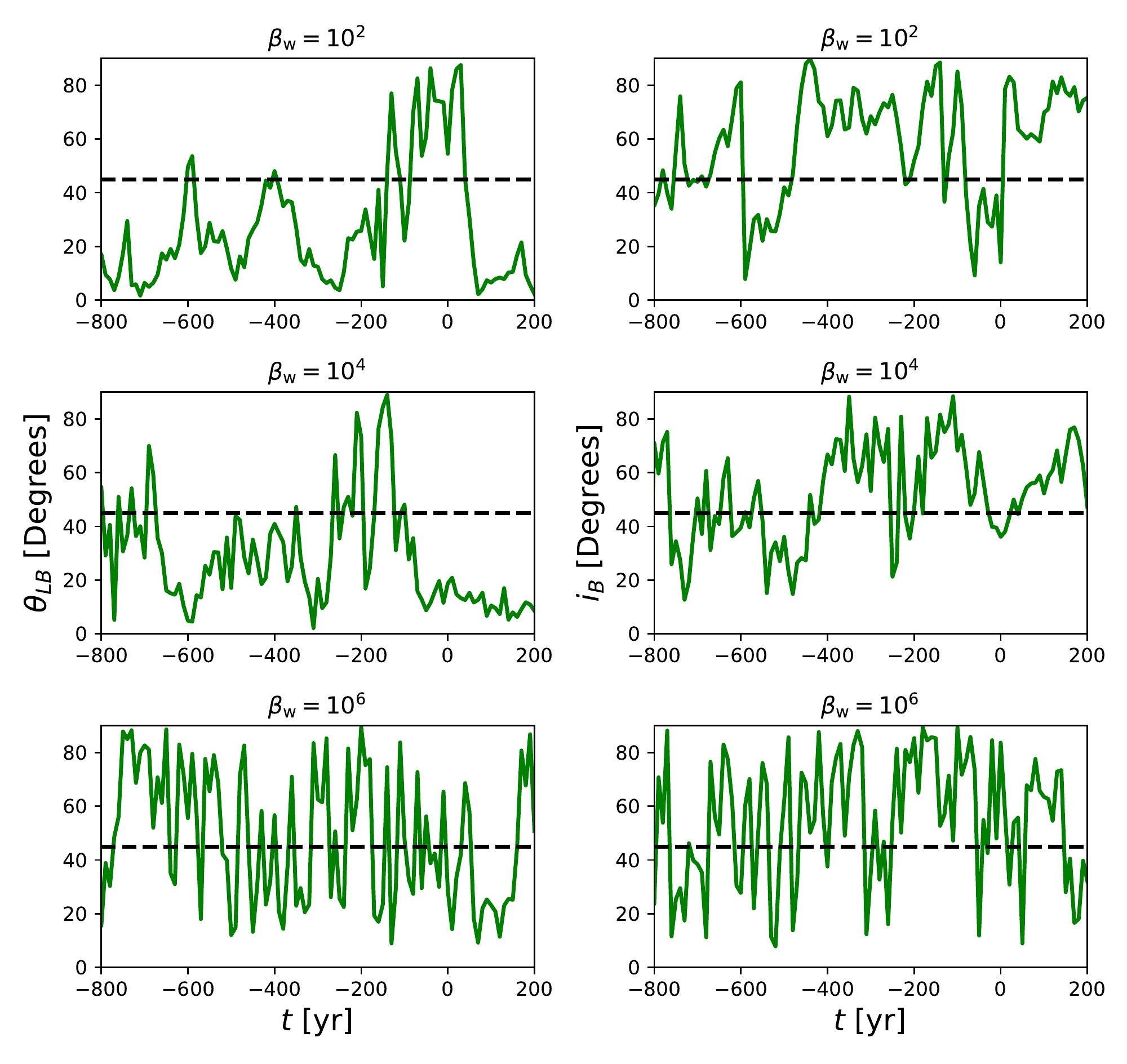}
\caption{Angles plotted vs. time in the $\beta_{\rm w} = 10^2,10^4,$ and $10^6$ (top, middle, and bottom panels, respectively) wind-fed MHD simulations of R20. Left column: Angle between $\langle \hat B\rangle$, the angle-averaged magnetic field direction, and $\langle \hat L\rangle$, the angle-averaged angular momentum direction,  $\theta_{LB}$.  Right column: the inclination angle of $\langle \hat B \rangle$ with respect to the line of sight, $i_B$ (right column).   These quantities are averaged over the innermost radii.  For reference, the dotted horizontal lines represent $45^\circ$.
The magnetic field that results from the more strongly magnetized winds (e.g., $\beta_{\rm w} = 10^2,10^4$) is more likely to be aligned with the angular momentum direction of the gas because it is strong enough to maintain its initial coherence, with $\theta_{LB} \lesssim 30^\circ$ most of the time.   The field resulting from more weakly magnetized winds (e.g., $\beta_{\rm w} = 10^6$), on the other hand, is essentially uncorrelated with the angular momentum direction because it easily gets tangled by the stochastic motion of the flow.  $i_B$ varies from 0$^\circ$--90$^\circ$ in all three simulations.  Compared to the $\beta_{\rm w}=10^6$ field, which oscillates rapidly in time with no preferred inclination, the $\beta_{\rm w}=10^2$ and $\beta_{\rm w}=10^4$ fields tend to be preferentially closer to edge-on ($90^\circ$), though they still show instances of being nearly face-on ($0^\circ$).  This demonstrates the difficulty in predicting the horizon-scale counterpart of $i_B$.  } 
\label{fig:lb_angle}
\end{figure*}

Defining $\langle \hat B\rangle  \equiv \langle \mathbf{ B} \rangle/|\langle\mathbf{B}\rangle|$ and $\langle \hat L \rangle \equiv \langle \rho \mathbf{l}\rangle/|\langle \rho \mathbf{l}\rangle|$, where $\mathbf{l} = \mathbf{r} \times \mathbf{v}$ and $\langle \rangle$ denotes an average over all angles, the angle between these two vectors is $\theta_{LB} \equiv \arccos \left(\left|\langle \hat L \rangle \cdot \langle \hat B \rangle\right|\right)$.  As discussed in the main text, $\theta_{LB}$ is a proxy for the resulting tilt of the horizon-scale accretion flow with respect to the angular momentum of the gas at large radii.  It is shown vs. time in the left column of Figure \ref{fig:lb_angle} for the $\beta_{\rm w} = 10^2,10^4$, and $10^6$ R20 simulations.  Note that $\langle \hat B \rangle$ and $\langle \hat L \rangle$ were also radially averaged over ($5 \times 10^{−4}$ pc, $3 \times 10^{-2}$ pc) before computing $\theta_{LB}$.  
For $\beta_{\rm w} = 10^{2}$ and $\beta_{\rm w}=10^4$, $\theta_{LB}$ is generally small, $\lesssim$ 30$^\circ$ at most times.  This is because 1) the components of the field initially perpendicular to the angular momentum vector contribute mainly to the resulting toroidal field which averages out over angle and 2) the field in these simulations is dynamically important for all radii $\lessapprox 10^{-2}$--$10^{-1}$ pc (Figure 6 in R20), so that it is able to resist the motion of the gas and retain its component initially parallel to the angular momentum vector.      For the $\beta_{\rm w}=10^6$ simulation, however, $\theta_{LB}$ oscillates rapidly in time about $45^\circ$ and has no preferred values.  The field strength in this simulation is never dynamically important across the R20 domain and thus the initial vertical component is free to be tangled incoherently by the motion of the gas, leading to an essentially random net magnetic field direction.  

\begin{figure}
\includegraphics[width=0.45\textwidth]{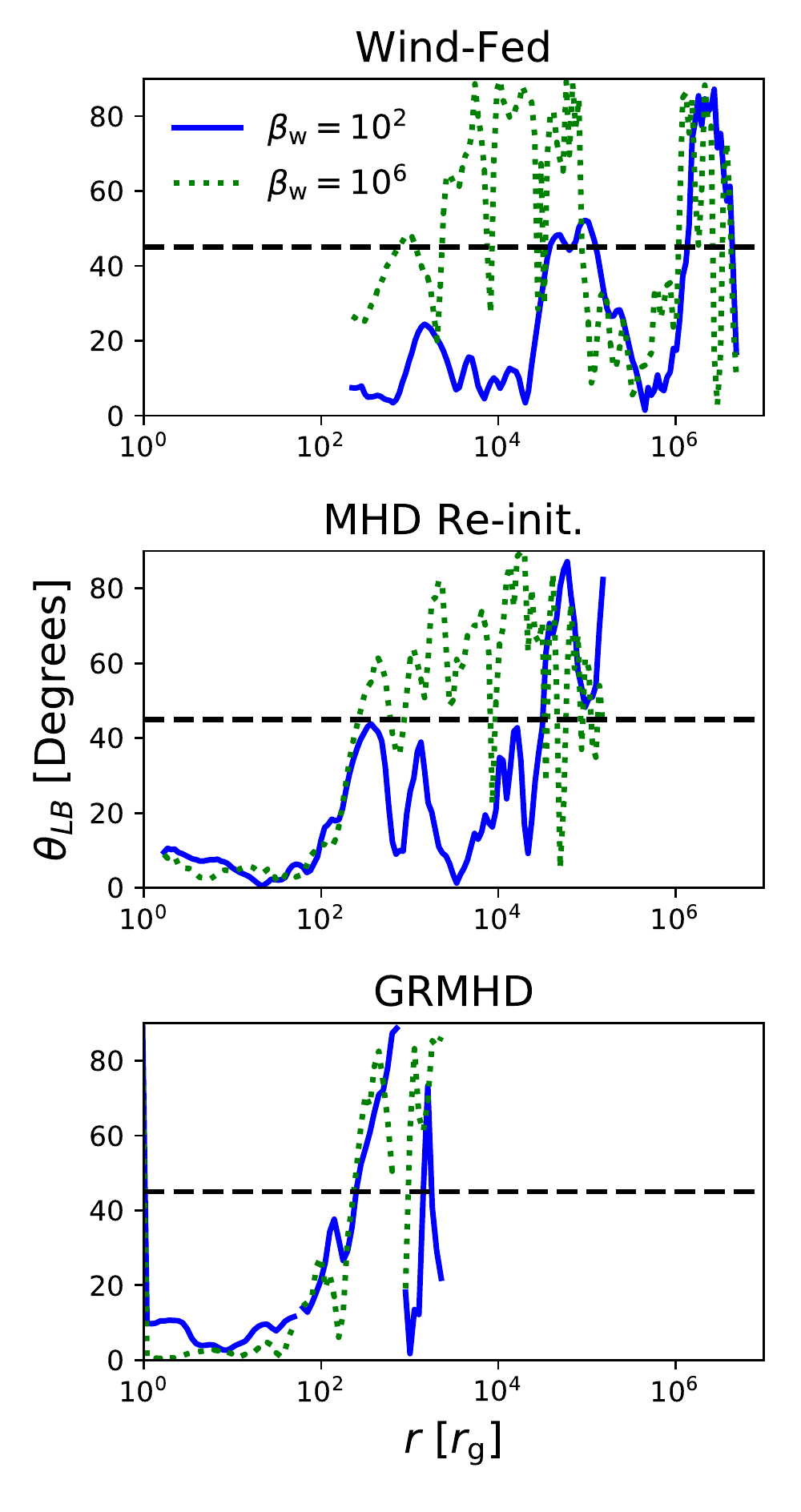}
\caption{$\theta_{LB}$ vs. radius where $\theta_{LB}$ is the angle between $\langle \hat B\rangle$, the angle-averaged magnetic field direction, and $\langle \hat L\rangle$, the angle-averaged angular momentum direction.  Solid lines are $\beta_{\rm w}=10^2$ and dotted lines are $\beta_{\rm w}=10^6$ for our MHD wind-fed simulations (top), our re-initialized intermediate scale MHD simulations (middle), and our GRMHD simulations (bottom).  The dashed horizontal lines indicate 45$^\circ$.   Magnetic flux that builds up near the inner boundaries forces the angular momentum and magnetic field direction to align ( i.e.,  $\theta_{LB} \lesssim 10^\circ$) in the inner $r\lesssim 100  r_{\rm in}$, where $r_{\rm in}$ is the inner boundary radius (or the event horizon radius in GR).  This happens as long as $\beta$ is sufficiently small, $\lesssim 10$, which is the case for all of our $\beta_{\rm w}=10^2$ simulations and the two smaller scale $\beta_{\rm w}=10^6$ simulations.     } 
\label{fig:lb_angle_radius}
\end{figure}

Figure \ref{fig:lb_angle_radius} demonstrates the alignment of $\langle \hat L \rangle$ and $\langle \hat B \rangle$ near the inner boundaries of our simulations by plotting $\theta_{LB}$ vs. radius.  For $\beta_{\rm w} = 10^2$, the magnetic field is sufficiently strong in the wind-fed MHD, the intermediate scale MHD, and the GRMHD simulations to tilt the angular momentum direction of the gas in the inner $\sim$ 100 $r_{\rm in }$ of the domain, where $r_{\rm in}$ is the inner boundary radius (or event horizon radius).   For the wind-fed MHD case, this behavior occurs at artificially large radii because of the larger $r_{\rm in}$.   Thus, the original R20 $\theta_{LB}$ between $\sim$ $10^2$--$10^4 r_{\rm g}$ is ``forgotten'' in the two smaller scale simulations, in which $\theta_{LB}$ between $\sim$ $10^2$--$10^4 r_{\rm g}$  more naturally connects to the $r \gtrsim 10^4 r_{\rm g}$ curve in the R20 simulation.  The alignment between $\langle \hat L \rangle$ and $\langle \hat B \rangle$ near the inner boundary is seen also for the two smaller scale $\beta_{\rm w}=10^6$ simulations but not in the wind-fed ,$\beta_{\rm w}=10^6$ simulation at larger scales where the field is too weak to sufficiently torque the gas.  

Plotted in the right column of Figure \ref{fig:lb_angle} is the inclination angle of the net magnetic field with respect to the line of sight, $i_{B} \equiv \arccos\left(\left| \langle \hat B \rangle_z \right|\right)$, a rough proxy for the inclination angle of the ultimate horizon-scale angular momentum.  Because the clockwise stellar disk has an inclination angle of $\sim$ $53^\circ$ \citep{Belo2006} and as just described $\theta_{LB}$ tends to be small for $\beta_{\rm w}=10^2$ and $\beta_{\rm w}=10^4$, these simulations show $i_B \gtrsim 60^\circ$, that is, nearly edge-on inclinations most of the time.    Both, however, have instances where $i_B \lesssim 30^\circ$ and is thus closer to face-on. $i_B$ in the $\beta_{\rm w} = 10^6$ simulation oscillates rapidly with no clear preference for a face-on or edge-on inclination.  
Since $\beta_{\rm w}$ is unknown and the detailed behavior of the curves in Figure \ref{fig:lb_angle} is moderately sensitive to the precise details of the R20 simulations (e.g., the spin axis of the stars and the inner boundary radius), we cannot make a robust prediction for the inclination angle of the horizon-scale accretion flow surrounding Sgr A*.

\end{document}